\documentclass[prl, reprint, nofootinbib, showpacs,preprintnumbers,amsmath,amssymb]{revtex4}
\usepackage{varioref,exscale,latexsym,amsmath,amssymb}
\usepackage{graphicx}

\usepackage{graphicx}
\usepackage{slashed}
\usepackage{dcolumn}
\usepackage{bm}
\usepackage{hyperref}
\usepackage{color}
\usepackage{xcolor}

\newcommand{\beq}{\begin{equation}}
\newcommand{\eeq}{\end{equation}}
\newcommand{\bea}{\begin{eqnarray}}
\newcommand{\eea}{\end{eqnarray}}

\def\lsi{\raise0.3ex\hbox{$<$\kern-0.75em\raise-1.1ex\hbox{$\sim$}}}
\def\gsi{\raise0.3ex\hbox{$>$\kern-0.75em\raise-1.1ex\hbox{$\sim$}}}

\def\beq{\begin{equation}}

\def\eeq{\end{equation}}

\def\beqa{\begin{eqnarray}}

\def\eeqa{\end{eqnarray}}

\begin{document}
\preprint{ACFI-T21-14}

\title{{\bf On Quadratic Gravity}}

\medskip\

\medskip\

\author{John F. Donoghue${}^{1}$}
\email{donoghue@physics.umass.edu}
\author{Gabriel Menezes${}^{2}$}
\email{gabrielmenezes@ufrrj.br}

\affiliation{
${}^1$Department of Physics,
University of Massachusetts,
Amherst, MA  01003, USA\\
${}^2$Departamento de F\'{i}sica, Universidade Federal Rural do Rio de Janeiro, 23897-000, Serop\'{e}dica, RJ, Brazil}

\begin{abstract}
We provide a brief overview of what is known about Quadratic Gravity, which includes terms quadratic in the curvatures in the fundamental action. This is proposed as a renormalizeable UV completion for quantum gravity which continues to use the metric as the fundamental dynamical variable. However, there are unusual field-theoretic consequences because the propagators contain quartic momentum dependence. At the present stage of our understanding, 
Quadratic Gravity continues to be a viable candidate for a theory of quantum gravity. 
\end{abstract}
\maketitle

\section{Introduction}
Quadratic gravity is the theory defined by the action\footnote{One can add a cosmological constant of course but that does not appear to play a significant role in the analysis of the theory, so we will not include it here.}
\beq\label{basicaction}
S_{\textrm{quad}} = \int d^4x \sqrt{-g}
\left[\frac{2}{\kappa^2}R+\frac{1}{6 f_0^2} R^2 - \frac{1}{2\xi^2} C_{\mu\nu\alpha\beta}C^{\mu\nu\alpha\beta}\right]
\eeq
where $\kappa^2 =32\pi G$, $f_0,~\xi$ are dimensionless coupling constants and $C_{\mu\nu\alpha\beta}$ is the Weyl tensor. We here ignore surface terms and can also write
\beq \label{weylidentity}
 \int d^4x \sqrt{-g}~\left( -\frac{1}{2\xi^2} C_{\mu\nu\alpha\beta}C^{\mu\nu\alpha\beta} \right)
 =  \int d^4x \sqrt{-g}~ \left[-\frac{1}{\xi^2}\left(R_{\mu\nu}R^{\mu\nu}
- \frac13 R^2\right) \right] \ \ .
\eeq
In some ways this is the most conservative UV complete quantum theory of gravity in that it continues to use the metric as the fundamental field variable and is renormalizeable  \cite{Stelle:1976gc, Tomboulis, Shapiro, Salvio:2018crh, Strumia, Narain, Donoghue:2018izj, Einhorn, Anselmi, Mannheim, Holdom, unitarity}. It is often overlooked in discussions of quantum gravity in favor of options which are very much more exotic. The reason for this (to be discussed more fully below) is that it must break at some aspect of our usual formulation of quantum field theory. However there is often not agreement on what exactly the problem is. 

The relevant literature extends over many decades and is somewhat hard to penetrate. There has been renewed modern interest in the topic. The purpose of the present paper is to provide an overview of the issues, with of course an emphasis on the work which we have been doing\footnote{This paper is a distant reflection of a talk presented by JFD at the on-line workshop Quantum Gravity, Higher Derivatives and Non-locality, March 2021, and is to be published in a special volume on this topic.} \cite{Donoghue:2018izj, unitarity, Donoghue:2017vvl, Donoghue:2018lmc, Donoghue:2019ecz, Donoghue:2020mdd,  Donoghue:2021eto,  Donoghue:2021meq}. While some phenomenology has been explored \cite{Holdom2, Stelle2}, the emphasis here will be on whether Quadratic Gravity works as a viable quantum field theory, and what its novel features are.  Because of space limitations, the explanations and referencing are somewhat abbreviated - more of both can be found in our cited papers. 

\section{The good news}

The reason that quadratic gravity is renormalizeable is that the propagators fall as $1/q^4$ in the UV. The curvature is second order in derivatives and so the terms which are quadratic in the curvature are then fourth order in derivatives. This UV behavior partially  damps the high energy divergences, turning them into renormalizeable ones. The power counting is done in a more general context in \cite{Donoghue:2019clr}. But it can be seen in a simple way by observing that the most divergent vertices in the UV come from the curvature squared terms, and noting that the coefficients of these,  $f_0$ and $\xi$, are dimensionless. Dimensional analysis then tells us that these terms will not produce the inverse powers of masses needed to form higher dimensional operators.

The coupling constant $\xi^2$ is asymptotically free  and $f_0^2$ is not, when used with the signs of Eq. \ref{basicaction}. Early papers had $f_0^2$ appearing with a different sign in the action, and then it appeared to be asymptotically free \cite{Julve:1978xn,Fradkin:1981hx}. However, this leads to a high mass spin-zero tachyon  \cite{Strumia} - that is, a pole at spacelike\footnote{Our metric is (+,-,-,-).} $q^2$. The signs of Eq. \ref{basicaction} are chosen to avoid tachyons. 

\section{The bad news}

K\"{a}llen and Lehmann \cite{Kallen:1952zz,Lehmann:1954xi} have told us that propagators have a spectral representation
\beq
D(q)=  \frac{1}{\pi}\int_{0}^\infty ds \frac{\rho(s)}{q^2-s+i\epsilon}
\eeq
with the spectral density $\rho(s) $ being positive definite. The ingredients to this are all properties which we expect in a quantum field theory: unitarity, causality, positive norm states, Lorentz invariance, etc. 
The spectral representation tells us that propagators cannot fall faster than $1/q^2$ at large $q^2$, But in quadratic gravity, propagators do fall faster than this, going as $1/q^4$. So something has to give - we can't have all the QFT properties which we normally expect. 

There is a caveat to this argument. The K\"allen-Lehmann argument does not apply to propagators in gauge theories. Indeed the gluon propagator in Landau gauge falls logarithmically faster than $1/q^2$ \cite{Oehme:1979ai}. The exception comes because of the unphysical states which appear when attempting a covariant description of gauge bosons, which however are gauge artifacts. This exception is relevant to spin-zero part of the graviton propagator in quadratic gravity. In this case there is a negative norm massless pole in covariant gauges, just as there is in normal General Relativity \cite{Alvarez-Gaume}. This is a gauge artifact. One might hope that the caveat also saves the odd behavior in the spin-two part of the propagator from being problematic. However, it does not seem to do so. The spin-two propagator enters into diagrams in the same way as non-gauge propagators do in other higher derivative theories, and seems to have the same behavior when analyzed. 

The problem can be seen directly. The bare spin-two propagator has the form
\beq\label{partialfraction}
iD_2(q) = \frac{i}{q^2-\frac{q^4}{M^2}} = \frac{i}{q^2} - \frac{i}{q^2-M^2}
\eeq
with $M^2= 2\xi^2/\kappa^2 $. The relative minus sign is required to get the asymptotic $1/q^4$ behavior. But of course it is the minus sign which signals the problem.  However, this problem could perhaps be manifest in various different ways related to unitarity, stability, causality etc. Further work is needed to sort out the implications. 
The interpretation that emerges is that causality is the broken ingredient, and that the extra minus sign in the propagator signals a particle whose propagation is time-reversed from usual particles. 

The exploration of higher derivative field theories goes back to the work by Lee and Wick in the 1960's \cite{Lee:1969fy, Lee:1969fz,  Lee:1970iw, Coleman, Cutkosky:1969fq,  Grinstein:2007mp}. They proposed a higher derivative version of QED as a way to make the theory finite. While our motivation here is different, many of the field-theoretic aspects can be traced back to the Lee-Wick program.

This is probably a good place to point out that it is not clear that all of our traditional techniques need to produce the same outcome when applied to higher derivative theories. The equivalences of Hamiltonian and Lagrangian methods, of canonical and path integral quantization, and of Euclidean and Lorentzian formulations, have been developed with standard theories and may have modifications when applied to higher derivative theories. Our treatment starts from the Lorentzian quantum path integral over the field variable with a higher derivative Lagrangian, including interactions from the start. 

\section{Ostrogradsky}

Theories with higher time derivatives are often rejected immediately because of the analysis by Ostrogradsky that says that the classical theory has an instability, with a Hamiltonian which is not positive definite \cite{ Ostrogradsky:1850fid, Woodard:2015zca}. However, a classical instability need not be a quantum one - the Dirac Hamiltonian is a counter-example. In fact, the path integral treatment of a higher derivative theory differs from Ostrogradsky's construction. In at least some cases, this can avoid the Ostrogradsky instability. This section is a mini-summary of our paper on this topic using a very simple model to demonstrate this \cite{Donoghue:2021eto}. 

Ostrogradsky's construction is designed to find a Hamiltonian whose use in Hamilton's equations reproduces the Euler-Lagrange equations, which have been derived from the initial Lagrangian. Theories with higher derivatives have extra degrees of freedom. Consider for example a theory with a Lagrangian
\beq
{\cal L}(\phi, \chi) = \frac12 \partial_\mu \phi \partial^\mu \phi - \frac1{2M^2} \Box \phi \Box \phi - g \phi \chi^\dagger \chi  \ \  .
\eeq
Here, $\chi$ represents some scalar field with a normal Lagrangian. In this case, Ostrogradsky's construction chooses the canonical coordinates to be
\beq\label{coordinates}
\phi_1 = \phi  ~~~~~,~~~~~
\phi_2 = \dot{\phi}   \  \ . 
\eeq
Following usual procedures then leads to the Hamiltonian
\beq\label{Hamiltonian}
{\cal H} = \pi_1 \phi_2 + \pi_2 \left(\nabla^2 \phi -M^2\pi_2  \right) - {\cal L}(\phi_1,\phi_2, \nabla^2 \phi -M^2\pi_2 ) \ \ ,
\eeq
where $\pi_1,~\pi_2$ are the canonical momenta conjugate to $\phi_1,~\phi_2$. The instability is visible in the first term, as $\pi_1$ can have either sign and there is no compensating factor of $\pi_1$ in the rest of the Hamiltonian. 

However, the choice of canonical coordinates of Eq. \ref{coordinates} seems odd for the quantum theory. As $M^2\to \infty$ one does not recover the normal result. There is not a perturbative limit for large $M^2$ as would be used in an effective field theory. Moreover, there is not a firm reason why Hamilton's equations in their usual form must be satisfied in the higher derivative quantum theory. The Hamiltonian is used in quantum physics, along with the quantization procedure, to identify energy eigenstates - which is a different objective. In this case, there is a different outcome.

Path integral quantization starts with 
\beq
Z_\phi [\chi] =\int [d\phi] e^{i \int d^4 x ~ {\cal L}(\phi, \chi) }
\eeq
We can introduce an auxiliary field $\eta$ which, when you integrate it out, reproduces the initial Lagrangian. This is
\beq
{\cal L}(\phi, \eta) =  \frac12 \partial_\mu \phi \partial^\mu \phi  - \eta \Box \phi + \frac12 M^2 \eta^2 -g\phi \chi^\dagger \chi \ \ .
\eeq
This results in
\beq
Z_\phi [\chi] = \int [d\phi ] [d\eta] e^{i \int d^4x [{\cal L}(\phi,\eta)]} \ \ .
\eeq
Then we redefine the field using $\phi(x) = a(x) -\eta(x)$ and obtain
\beqa\label{jointPI}
Z_\phi [\chi] &=& \int [da] e^{i \int d^4 x \left[\frac12 \partial_\mu a \partial^\mu a -g a \chi^\dagger \chi\right] } \nonumber \\
  &\times& \int [d\eta] e^{-i \int d^4 x \left[\frac12 \partial_\mu \eta \partial^\mu \eta -\frac12 M^2 \eta^2 -g \eta \chi^\dagger \chi \right]}  \nonumber \\
  &=& Z_a \times Z_\eta
\eeqa
without any approximations. 

The second path integral $Z_\eta$ is actually well defined despite the unusual overall factor of $-i$. It is just the complex conjugate of our usual Gaussian path integral. One obtains
\beq
Z_\eta = N e^{\int d^4x d^4y \frac12 g\chi^\dagger(x)\chi(x) ~i D_{-F}(x-y) ~g\chi^\dagger (y)\chi(y)} \ \ .
\eeq
with
\beq
iD_{-F}(x-y) = \int \frac{d^4k}{(2\pi)^4} \frac{-i}{k^2- M^2-i\epsilon} e^{- i k \cdot (x-y)} \ \ .
\eeq
being the complex conjugate of the usual result.
If $M$ is very large, the propagator becomes local and  just leaves a contact interaction $(\chi^\dagger \chi)^2 $ which does not destabilize the theory if such an interaction appeared in the original theory. This satisfies the principle of decoupling -- as $M^2\to \infty$ we should get only shifts in the parameters of the low energy Lagrangian and effects suppressed by powers of $M^2$. 

We commonly refer to the classical limit as that of $\hbar\to 0$. But this is not really correct as $\hbar$ is a fixed number. The classical limit is obtained by using those kinematic and spatial conditions where $\hbar$ is unimportant.
The classical limit of this theory is at low energy and long wavelength where quantum effects are not important. The degrees of freedom here are $a$ and $\chi$ with normal Lagrangians. The theory does not exhibit the Ostrogradsky instability.

In canonical quantization, the procedure looks different than when using path integrals. While one can use the same field redefinitions, one must also modify the quantization rules in order to obtain positive energy eigenstates \cite{Lee:1969fy,   Salvio:2015gsi, Raidal:2016wop}. Very roughly, this means using $[a,a^\dagger]=-1$ instead of the usual rule. But the main point, in both path integral and canonical quantization, is that the Ostrogradsky construction is not that of the quantum theory.

The spin-two sector of quadratic gravity appears similar to this example. The interactions are more complicated of course and the theory does not perfectly factorize. But the path integral has a similar form, and a similar classical limit. 

\section{The spectrum}

Another misconception about this class of theories is the idea that the massive ghost carries negative energy. Again, a classical analysis would seem to show this. (For example the unusual minus sign in $Z_\eta$ of Eq. \ref{jointPI} could be interpreted as a negative Lagrangian, leading to a negative Hamiltonian.) As mentioned, canonical quantization can be modified to lead to positive energies. One can see the same in a path integral by studying the production of the massive state, which also yields more information about the nature of the resonance. 

The heavy particle in the spin-2 graviton propagator can be produced by the scattering of matter particles, generically $\phi +\phi \to R \to \phi +\phi$ , where here $\phi$ is a generic matter particle (or massless graviton) . This coupling to matter also gives the graviton a self-energy - the vacuum polarization diagram - which contains an imaginary part for timelike momenta above threshold. The propagator in the relevant region has the form \cite{Donoghue:2018izj}
\beq\label{spintwo}
iD_2(q) = \frac{ i}{\left\{ {q^2+i\epsilon}- \frac{\kappa^2 q^4}{2\xi^2(\mu)}
- \frac{\kappa^2 q^4 N_{\textrm{eff}}}{640\pi^2} \left[\ln \left(\frac{|q^2|}{\mu^2}\right)-i\pi \theta(q^2)\right]
\right\}}
\eeq
and with $N_{\rm eff} $ a number which counts the effective number of light degrees of freedom. The corresponding spin-two scattering amplitude \cite{unitarity, Han} is
\beq\label{t2amplitude}
T_2(s) = - \frac{N_{\textrm{eff}} s}{640 \pi}\,{D}_2(s)  \ \  .
\eeq

With the correct choice of the sign of the coupling $\xi^2$, the resonance happens for timelike momenta. Since the initial and final $\phi+\phi$ states carry positive energy, the resonance also has positive energy. However, the form of the resonance is unusual in two respects. We can see this by looking at the propagator near the resonance. Expanding  Eq. \ref{spintwo} near $q^2\sim M^2$, with $M^2$ being the real part of the pole location, one finds the form
\beq
iD_2\sim \frac{-i}{q^2 -M^2 -i\gamma}
\eeq
with $\gamma>0$. There is the unusual minus sign in the numerator, which is however expected as already explained in Eq. \ref{partialfraction}. In addition, the imaginary part of the denominator comes with the opposite sign from a usual resonance. Basically, the result is the complex conjugate of a usual resonance propagator. This is a characteristic feature of this class of higher derivative theories. The combination of the two unusual signs means that the imaginary part of $D_2$ is the same as a normal resonance - a feature which is important for unitarity. 

Despite the change in sign in the width, the heavy particle exhibits exponential decay rather than exponential growth. This can be seen by using the time ordered form of the propagator after performing the $q_0$ integration,
\beq
D_2(t, \vec{x}) = \Theta (t)D_{\textrm{for}} (x) + \Theta (-t) D_{\textrm{back}}(x)
\eeq
with 
\beq\label{forward}
D_{\textrm{for}}(t,\vec{x}) = -i\int \frac{d^3q}{(2\pi)^3 }\left[\frac{e^{-i(\omega_q t -\vec{q}\cdot\vec{x})}}{2\omega_q } - \frac{e^{i(E_q t -\vec{q}\cdot\vec{x})}}{2(E_q +i\frac{\gamma}{2E_q})} e^{-\frac{\gamma t}{2E_q}}\right]
\eeq
and 
\beq\label{backwards}
D_{\textrm{back}}(t,\vec{x}) = -i\int \frac{d^3q}{(2\pi)^3 }\left[\frac{e^{i(\omega_q t -\vec{q}\cdot\vec{x})}}{2\omega_q } - \frac{e^{-i(E_q t -\vec{q}\cdot\vec{x})}}{2(E_q +i\frac{\gamma}{2E_q})} e^{-\frac{\gamma |t|}{2E_q}}\right]. \ \ .
\eeq
Here $\omega_q=|\mathbf{q}|$ for the massless graviton, and $E_q=\sqrt{\mathbf{q}^2+M^2}$ for the massive resonance.

\section{Causality}

How shall we interpret the spectrum described above? The major clues are 1) that the heavy particle behaves as if it were formed from a path integral using $e^{-iS} $ instead of $e^{+iS}$ and 2) that the resulting propagators are complex conjugates of normal propagators. These make sense when we remember that time-reversal symmetry is anti-unitary, i.e. involving complex conjugation. Both the path integral and the resonance propagator are the time reversed versions of the usual results \cite{Donoghue:2019ecz,  Donoghue:2020mdd}. This interpretation can be confirmed by looking at the time ordered propagators of Eqs. \ref{forward} and \ref{backwards}. Normal particles propagate positive energy forward in time, but the heavy resonance propagates positive energy backwards in time.

It is often not stated, but QFT comes with an {\em arrow of causality }  \cite{Donoghue:2019ecz, Donoghue:2020mdd}, which differentiates the past lightcone from the future lightcone. It is contained in the $+i\epsilon$ in the Feynman propagator, which can be traced back to the use of $e^{+iS}$ in the path integral (or to the $+i\hbar$ in canonical commutators). This feature specifies the time direction in which positive energy reactions proceed\footnote{While it is not related to quadratic gravity, we have also pointed out how this feature explains the arrow of thermodynamics \cite{ Donoghue:2020mdd}}. 

The interpretation is then of a time-reversed unstable particle. Generally we refer to a propagator with a minus sign in the numerator as a ghost. However, Fadeev-Popov ghosts come with the usual $+i\epsilon$ in the denominator. 
Our present case is different because of the opposite sign in the denominator. We have proposed using the phrase {\em Merlin modes} (after the wizard in the Arthurian tales who ages backwards in time) to emphasize this distinction. 

Here is a place where the usual rules of QFT are no longer fulfilled. Causality and analyticity of amplitudes are tightly tied together, and the analyticity property of Feynman diagrams has been changed. There are now dueling arrows of causality, with the massless graviton and matter fields carrying the usual directionality, and the heavy Merlin mode carrying the other. This leads to a violation of causality on scales over which the Merlin mode propagates. While there are in principle signals of this behavior \cite{Coleman, Grinstein:2007mp, Alvarez:2009af}, for quadratic gravity they are proportional to the Planck time scale, which is far too small to be observable. 

This feature explains the failure of the K\"allen-Lehmann representation in higher derivative theories. Coleman has noted that in this case it is replaced by a different spectral representation \cite{Coleman}
\begin{equation}\label{Coleman}
D(q) = \frac{1}{q^2+i\epsilon} - \frac{\beta}{q^2-M_r^2} - \frac{\beta^*}{q^2-M_r^{*2}} + \frac{1}{\pi}\int_{0}^\infty ds \frac{\rho(s)}{q^2-s+i\epsilon}
\end{equation}
where here $M_r^2$ and $M_r^{*2}$ are complex pole locations, $\beta, ~\beta^*$ are the residues and $\rho(s)$ is a positive definite spectral function. Note the extra pair of poles which are complex conjugates of each other. This representation allows the $1/q^4$ fall off at large $q^2$ through cancellations between the various ingredients. The early literature focused on the pair of complex poles, but there is a third resonance structure in $\rho(s)$ which is a Breit-Wigner-like form \cite{ Grinstein:2007mp, Donoghue:2018lmc}.  The pole in $\rho(s)$ compensates one of the complex conjugate poles. We then have two descriptions of the propagator - the original form with one pole and the Coleman representation with three.

\section{Unitarity}

One naturally might worry also about whether unitarity is satisfied in such theories. After all, there is a negative norm state. Does this appear in the unitarity relation? That would seem to mess up unitarity. 

Actually unitarity does work. We have given a formal proof \cite{unitarity}, which is somewhat dense as it follows Veltman's largest-time method \cite{Veltman:1963}. But the basic idea is simple. Even with normal unstable particles, you are not supposed to include the unstable particle in the unitarity relation. Only the stable particles which appear as asymptotic states are to be included. This makes sense as the $S$ matrix deals only with asymptotic in and out states. But it also seems odd because we are used to dealing with weakly decaying particles, such as the pion for example, as if they were stable and studying unitarity without considering their decay products. Nevertheless, the right answer is to only consider fully stable particles in the unitarity relation. This is classic work by Veltman \cite{Veltman:1963}. Our intuition works in practice because one can show that in the narrow width approximation - when the width is very small - the discontinuity of the stable particles in the initial and final states becomes the same as if we had treated the resonance as if it were stable \cite{unitarity}. 

For the Merlin modes we see the same result. Only the normal decay products count in the unitarity relation. There are no negative norm asymptotic states. Because all of the in and out states have their normal discontinuities, unitarity is satisfied. 

The direct production of the the Merlin resonance in quadratic gravity described above is the simplest example. The  spin-two scattering amplitude of Eq. \ref{t2amplitude} has the form
\beq
T_2(s) = \frac{A(s)}{f(s) -i A(s)}= \frac{A(s)[f(s)+i A(s)]}{f^2(s)+A^2(s)}
\eeq
 with
\beq
A(s) =  - \frac{N_{\textrm{eff}} s}{640 \pi} \ \ .
\eeq
It readily checked that this form satisfies unitarity. 

However, there is a complication. The narrow width approximation does not work exactly the same way as with a normal resonance. In some loop diagrams, if one calculates the discontinuity using only the stable decay products, one obtains the usual unitarity relation, and this has a well defined narrow width limit. However, if one calculates these loop diagrams assuming that the Merlin resonance is exactly stable one gets a different result unless one uses a contour integration that encloses the Merlin pole in a certain way. This feature has been known for a long time and is referred to as the Lee-Wick contour \cite{Lee:1969fy}. In simple examples it can be implemented, but work is needed to understand it more fully. 

\section{Known Unknowns}

Despite the aspects which are presently understood, there are still facets of Quadratic Gravity which are not known. By modifying QFT even slightly by using quartic propagators, all aspects need to be re-thought. This makes the theory interesting to explore. 

We have seen that the Ostrogradsky instability can be avoided in this class of theories near flat space. However, stability at higher curvatures is not known. Perhaps when the background curvature becomes of order the mass of the Merlin ghost there could be an instability. If so, the endpoint of that instability is relevant. It could be a good feature if the endpoint was a state with smaller curvature, perhaps with radiated particles. It could be fatal if it implies a runaway to infinite curvature. It is also possible that some quantum processes could trigger an instability in a way that is not visible in the tree-level analysis described above. 

There also needs to be more work understanding the quantum field theory of Quadratic Gravity. We have mentioned the need to use the Lee-Wick contour in certain loop diagrams. There may be the need for further modifications in other diagrams. Cutkosky et al   \cite{Cutkosky:1969fq} made an initial exploration of Lee-Wick theories without reaching a satisfactory conclusion. However, we note that they were studying only the complex conjugate poles of Eq. \ref{Coleman} and did not include the pole in the spectral function, so this study needs to be revisited. Presumably, the principle is that the unitary relations, calculated directly, need to be reproduced by the Feynman diagrams. Alternately, one can perhaps employ unitarity to reproduce the full amplitudes, as in modern unitarity based techniques. One of us (GM) has been exploring this promising pathway \cite{GMunitarity, GMdouble}

Anselmi has been exploring the Feynman rules for a similar class of theories, and has introduced prescriptions which seem to work \cite{Anselmi}. There is in principle a difference between Anselmi's program and what we have been describing. He defines his theory as the Euclidean version and tries to find rules for the analytic continuation to Lorentzian. For us, the fundamental version is Lorentzian and the continuation to Euclidean is just a way of doing integrals, etc. This again highlights the point that our usual equivalence between Euclidean and Lorentzian theories comes from our treatment of normal theories and may be changed in theories with higher derivatives. 

It would also be interesting to simulate a related theory with higher derivatives on a lattice. This would be a Euclidean theory, but it could provide a non-perturbative insight into the stability of the theories. It is also possible that the quadratic terms in the curvature could provide a good regularizer in numerical studies which try to simulate quantum General Relativity, as these terms help improve the high energy behavior. Using these as a regulator could then also probe Quadratic Gravity.

\section{Summary}

In some ways, Quadratic Gravity is a very conservative approach to quantum gravity as it maintains the paradigm of renormalizeable field theories like the rest of the Standard Model, and it does not introduce new fields. But it is not totally benign. Some features of our normal QFTs must be reworked. 

We have seen that causality is violated on small scales. For a theory of quantum gravity this may actually be an expected outcome as it is hard to see how we can maintain our usual ideas of causality in a fluctuating spacetime \cite{Donoghue:2021meq}. So we feel that this is not a feature which disqualifies the theory. 

Further explorations are still needed, as discussed above. There still could be potentially fatal obstacles. However, several of the standard objections to this class of theories seem to not be correct. At our present understanding, Quadratic Gravity can be a potential UV completion for quantum gravity.

\section*{Acknowledgements} The work of JFD has been partially supported by the US National Science Foundation under grant NSF-PHY18-20675. The work of GM has been partially supported by  Conselho Nacional de Desenvolvimento Cient\'ifico e Tecnol\'ogico - CNPq under grant 307578/2015-1 (GM) and Funda\c{c}\~ao Carlos Chagas Filho de Amparo \`a Pesquisa do Estado do Rio de Janeiro - FAPERJ under grant E-26/202.725/2018.

 \end{document}